\documentclass[usenatbib,twocolumn]{mnras} 
\usepackage[T1]{fontenc}
\usepackage{ae,aecompl}
\usepackage{graphicx}
\usepackage{hyperref}
\usepackage{amsmath}
\usepackage{amssymb}
\usepackage{mathtools}
\usepackage{bm}
\usepackage{cleveref}
\usepackage{widetext}
\usepackage{stfloats}

\usepackage{subcaption, caption}

\usepackage{float}
\usepackage{ dsfont }
\usepackage{ upgreek }

\usepackage{color,xcolor}

\title{Exotic Baryons in Hot Neutron Stars}

\author[A.~Issifu et al.]{
     A.~Issifu$^1$ 
    \thanks{E-mail: ai@academico.ufpb.br},     
    K.~D.~Marquez$^1$,
    M.~R.~Pelicer$^1$,~and
    D.~P.~Menezes$^1$ 
    \\ 
    \\
    $^1$Departamento de F\'{\i}sica - CFM - Universidade Federal de Santa Catarina  Florian\'opolis - SC - CP. 476 - CEP 88.040 - 900 - Brazil
    \\    }

\date{\today}

\begin{document}
    %
    %
    \pagerange{\pageref{firstpage}--\pageref{lastpage}} \pubyear{2023}
    \maketitle
    \label{firstpage}

\begin{abstract}
We study the nuclear isentropic equation of state for a stellar matter composed of nucleons, hyperons, and $\Delta$-resonances. We investigate different snapshots of the evolution of a neutron star, from its birth as a lepton-rich protoneutron star in the aftermath of a supernova explosion to a lepton-poor regime when the star starts cooling to a catalyzed configuration. We use a relativistic model within the mean-field approximation to describe the hot stellar matter and adopt density-dependent couplings adjusted by the DDME2 parameterization. We use baryon-meson couplings for the spin-$1/2$ baryonic octet and spin-$3/2$ decuplet determined in a unified manner relying on $\text{SU}(6)$ and $\text{SU}(3)$ symmetry arguments. We observe that $\Lambda$ is the dominant exotic particle in the star at different entropies for both neutrino-free and neutrino-trapped stellar matter. For a fixed entropy, the inclusion of new particles (hyperons and/or delta resonances) in the stellar matter decreases the temperature. Also, an increase in entropy per baryon ($1\;\text{to}\; 2$) with decreasing lepton number density ($0.4\;\text{to}\; 0.2$) leads to an increase in stellar radii and a decrease in its mass due to neutrino diffusion. In the neutrino transparent matter, the radii decrease from entropy per baryon $2$ to $T\,=\,0$ without a significant change in stellar mass. 

\end{abstract}

\begin{keywords}
stars: neutron, protostars
\end{keywords}


\section{Introduction}

The equation of state (EoS) is an essential tool for studying strongly interacting matter and performing astrophysical simulations of compact objects and has already been exploited in several forms~\citep{Typel, Dutra}. 
However, the microscopic composition of compact objects is still an open problem, and its resolution requires an enhanced understanding of the dense region of the EoS, both to understand current data and also to {accommodate} new observational advancements. Notable among emerging events that require the application of the EoS are multimessenger observations of binary neutron star mergers, isolated X-ray pulsars, and radio pulsars. The major constraints imposed on the EoS to study these objects include $\beta$-equilibrium, charge neutrality, and lepton number conservation -- see~\cite{Baym, Menezes} for recent reviews and references therein.

A hot and dense proto-neutron star (PNS) is a neutrino-rich object formed during a core-collapse supernova explosion or in a binary neutron star merger. The PNS evolves through several processes, including heat transfer, neutrino diffusion, deleptonization, and entropy gradients. When the star emits enough radiation, its mass decreases and its temperature drops to a point where matter becomes neutrino transparent and continues cooling till it catalyzes into a cold neutron star \citep{Glendenning}. The neutrino signature at the later stages of the evolution is determined by microscopic properties such as the EoS and its composition, neutrino opacity, and other microphysical properties that impact neutrino diffusion and finite entropy systems ~\citep{Sedrakian, Roberts, Prakash, Janka}. The study of gravitational collapse and supernova explosions are essential astrophysical events due to their rich physics and diversity. For instance, the process involves all four known fundamental forces of nature, making it an ideal laboratory for physics on different lengths and time scales and a testbed for new phenomena. The process starts in a strong gravitational field. Neutrino emission and deleptonization are weak interaction properties, the thermodynamic properties are governed by electrodynamics and strong interactions, while the change in the composition of the stellar gas is governed by nuclear and weak interactions~\citep{Camelio, Fischer, Pons, Camelio1}.

In this study, we  analyze the temperature profile and mass-radius diagram of the isotropic, static, spherically symmetric hot star containing the spin-$1/2$ baryon octet and the non-strange $J^P = 3/2^+$ decuplet. We investigate the behavior of the EoS and the particle abundances in the evolution of a newly born PNS until it catalyzes. 
Several studies of cold neutron stars have been carried out within the framework of relativistic models within a mean-field approximation taking into account all of the spin-$1/2$ octet and/or $\Delta$-resonances using various meson-baryon coupling formalism at zero temperature~\citep{Marquez, Schurhoff, Drago, Li, Raduta, Li1, Ribes, Zhu}. Studies on PNS at finite temperature and fixed entropy considering heavy baryons have also been done in~\citet{Sedrakian, Malfatti}. At the same time, hadron-quark phase PNS is also studied in~\citet{Shao} under fixed entropy conditions. In this work, we aim to give an overview of a neutron star's evolution, from its birth as a lepton-rich proto-neutron star in the aftermath of a supernova explosion to its final stages, when the star cools to a catalyzed configuration.

Different non-nucleonic degrees of freedom are considered to be present in neutron star matter, depending on the model adopted. In most of the contemporary literature, the nucleons and hyperons (the entire spin-$1/2$ baryon octet) are taken as the standard constituents of such objects, including the baryons of the spin-$3/2$ decuplet (especially $\Delta$-resonances) proving also to be relevant in the latest years.
The presence of hyperons and $\Delta$-resonances in the neutron star matter composition generally softens the EoS, lowering the maximum mass of the star below the expected threshold of $\sim 2\, M_\odot$~\citep{Antoniadis}. {The recent measurement of the massive pulsar {PSR J0740+6620} by NICER \citep{Cromartie2019, Fonseca:2021wxt} of $M=2.072^{{}+0.067}_{{}-0.066}$  M$_{\odot}$ and $R=12.39^{{}+1.30}_{{}-0.98}$ km, at a confidence interval of 68\% \citep{Riley:2021pdl}, gives a well-defined mass-radius window that must be reached by the NS description. As the RMF model parameters are fitted to reproduce nuclear matter observables, these astrophysical observations can be addressed 
mainly by adjusting the baryon-meson couplings of the non-nucleonic constituents of the stellar matter \citep{Weissenborn, Miyatsu, Lopes, Lopes_new}.}

In this study, we use baryon-meson couplings recently determined using group theory in~\citet{Lopes1}, to study the evolution of a PNS from its birth when $S/n_B = 1$ with trapped neutrinos, neutrino diffusion stage $S/n_B =2$ few seconds of its birth, neutrino transparent stage for $S/n_B = 2$, and finally to the formation of a cold neutron star at $T = 0$. 

The work is organized as follows: In Sec.~\ref{NS1} we present the details of the relativistic model in the mean-field approximation and the required conditions necessary for thermodynamics applications. The section is divided into two subsections; in Sec.~\ref{SN2} we present the details of the equations of state and in Sec.~\ref{NS3} we present the necessary equilibrium conditions for supernova physics. The results and analyses are contained in Sec.~\ref{RA}, where we discuss the particle abundances, the EoS, the temperature profiles, and the mass-radius diagrams.  The final findings are in Sec.~\ref{Con}, where we summarize all the stages of the star's evolution.

\section{Neutron star matter at finite entropy}\label{NS1}

\subsection{Equation of State}\label{SN2}

The Lagrangian of the relativistic model in the mean field approximation used to describe the hadronic interactions is given by
\begin{equation}
     \mathcal{L}_{\rm RMF}= \mathcal{L}_{H}+ \mathcal{L}_{\Delta}+ \mathcal{L}_{\rm mesons}+ \mathcal{L}_{\rm leptons},
\end{equation}
where the Dirac-type Lagrangian for the $J^P=1/2^+$ baryon octet is given by
\begin{align}
 \mathcal{L}_{H}= {}& \sum_{b\in H}  \bar \psi_b \Big[  i \gamma^\mu\partial_\mu - \gamma^0  \big(g_{\omega b} \omega_0  +  g_{\phi b} \phi_0+ g_{\rho b} I_{3b} \rho_{03}  \big)\nonumber \\
 &- \Big( m_b- g_{\sigma b} \sigma_0 \Big)  \Big] \psi_b,
\end{align}
and the Rarita-Schwinger--type Lagrangian for the $J^P=3/2^+$ particles of baryon decuplet is given by
\begin{align}
        \mathcal{L}_{\Delta}={}& \sum_{d\in \Delta}\Bar{\psi}_{d\nu}\Big[\gamma^\mu i\partial_\mu- \gamma^0\left(g_{\omega d}\omega_0 + g_{\rho d} I_{3d} \rho_{03} \right) \nonumber\\&-\left(m_d-g_{\sigma d}\sigma_0 \right)\Big]\psi_{d\nu}.
\end{align}
We stress that spin-$3/2$ baryons are described by the Rarita-Schwinger Lagrangian density and that their vector-valued spinor has additional components when compared to the four components in the spin-$1/2$ Dirac spinors. However, as shown in \citet{DePaoli}, spin-$3/2$ equations of motion can be written compactly as the spin-$1/2$ ones in the RMF regime. 
The mesonic part of the Lagrangian is given by
\begin{align}
 \mathcal{L}_{\rm mesons}= - \frac{1}{2} m_\sigma^2 \sigma_0^2  +\frac{1}{2} m_\omega^2 \omega_0^2 \label{lagrangian} +\frac{1}{2} m_\phi^2 \phi_0^2 +\frac{1}{2} m_\rho^2 \rho_{03}^2 ,
\end{align}
{where the interaction mediators are the scalar meson $\sigma$, the vector mesons $\omega$ and $\phi$ (which carries hidden strangeness), both isoscalars  and the isovector-vector meson $\vec\rho$. The subscript '$0$' here indicates that the field equations are calculated in the mean field approximation. Finally,} the free
leptons are described by the 
 Dirac Lagrangian
\begin{equation}\label{l1}
    \mathcal{L}_{\rm leptons} = \sum_L\Bar{\psi}_L\left(i\gamma^\mu\partial_\mu-m_L\right)\psi_L
\end{equation}
where the summation runs over all leptons considered in each stage of the star evolution.
For cold stellar matter, the index $L$ runs over electron and muons $L\in(e,\,\mu)$ and their corresponding antiparticles with a degeneracy factor of $\gamma_L=2J_L+1=2$, {with $J_L$ the total angular momentum  of the leptons.} For a finite temperature 
and in the case of  fixed entropy and lepton number density, we consider only the electron and its neutrino, since muons only become relevant after the star becomes neutrino-free~\citep{Malfatti}. In this case, we consider the left-handed electron neutrino in the Standard Model with a degeneracy of $\gamma_L = 1$ for a complete study. 


\begin{table}
\begin{center}
\begin{tabular}{ c c c c c c c }
\hline
 meson($i$) & $m_i(\text{MeV})$ & $a_i$ & $b_i$ & $c_i$ & $d_i$ & $g_{i N} (n_0)$\\
 \hline
 $\sigma$ & 550.1238 & 1.3881 & 1.0943 & 1.7057 & 0.4421 & 10.5396 \\  
 $\omega$ & 783 & 1.3892 & 0.9240 & 1.4620 & 0.4775 & 13.0189  \\
 $\rho$ & 763 & 0.5647 & --- & --- & --- & 7.3672 \\
 \hline
\end{tabular}
\caption {DDME2 parameters.}
\label{T1}
\end{center}
\end{table}

We use the density-dependent parametrization known as DDME2~\citep{ddme2}, where the meson couplings are 
adjusted by the expression
{ \begin{equation}
    g_{i b} (n_B) = g_{ib} (n_0)a_i  \frac{1+b_i (\eta + d_i)^2}{1 +c_i (\eta + d_i)^2},
\end{equation}
for $i=\sigma, \omega, \phi$ and 
\begin{equation}
    g_{\rho b} (n_B) = g_{ib} (n_0) \exp\left[ - a_\rho \big( \eta -1 \big) \right],
\end{equation}}
for $i=\rho$, with $\eta =n_B/n_0$. The model parameters are fitted from experimental constraints of nuclear matter at or around the saturation density, namely the binding energy, compressibility modulus, symmetry energy, and its slope, and are shown in Table~\ref{T1}, considering the associated bulk properties of nuclear matter at saturation $n_0=0.152$~fm$^{-3}$ as of being $B/A = -16.4$~MeV, $K_0=251.9$~MeV, $J=32.3$~MeV, and $L= 51.3$~MeV, which are in good agreement with current constraints
~\citep{Dutra,ddme2,reed2021,lattimer23}.

The fitting of the model-free parameters is made considering the pure nucleonic matter, and to determine the meson couplings to hyperons and deltas we define the ratio of the baryon coupling to the nucleon one as $\chi_{ib}=g_{i b}/g_{i N}$. 
{One way to extend the model parameterization to other baryonic degrees of freedom is to use flavor $\text{SU}(3)$ symmetry arguments to fix the values of the couplings, a procedure well adopted in the literature as it gets rid of the huge arbitrariness of the previously used recipes  \citep[c.f.][]{Stancu}. On the other hand, \cite{Lopes1} calculated the baryon-meson vector coupling constants of the spin-$1/2$ baryonic octet, and for the first time, calculated that of the spin-$3/2$ decuplet, in a model-independent way,} using the potentials  $U_\Lambda =-28$~MeV, $U_\Sigma= 30$~MeV, $U_\Xi=-4$~MeV, and $U_\Delta\approx -98$~MeV to fix the  scalar couplings. The values of $\chi_{ib}$ are shown in Tab.~\ref{T2} and are equivalent for the choice of $\alpha_V=0.5$ in the free parameter of the baryon-meson coupling scheme. 
Please note that some of the $\chi_{\rho b}$ parameters are different from the ones reported in ~\citet{Lopes1} because the model presented in~\citet{Lopes1} does not involve the  isospin projections in the Lagrangian terms  unlike the one under consideration here.

\begin{table}
\begin{center}
\begin{tabular}{ c c c c c } 
\hline
 b & $\chi_{\omega b}$ & $\chi_{\sigma b}$ & $\chi_{\rho b}$ & $\chi_{\phi b}$  \\
 \hline
 $\Lambda$ & 0.714 & 0.650 & 0 & -0.808  \\  
$\Sigma^0$ & 1 & 0.735 & 0 & -0.404  \\
  $\Sigma^{-}$, $\Sigma^{+}$ & 1 & 0.735 & 0.5 & -0.404  \\
$\Xi^-$, $\Xi^0$  & 0.571 & 0.476 & 0 & -0.606 \\
  $\Delta^-$, $\Delta^0$, $\Delta^+$, $\Delta^{++}$   & 1.285 & 1.283 & 1 & 0  \\
  \hline
\end{tabular}
\caption {The ratio of the baryon coupling to the corresponding nucleon coupling for hyperons and $\Delta$s.}
\label{T2}
\end{center}
\end{table}

From the Lagrangian, thermodynamic quantities can be calculated. The density of a baryons $b$ is given by
\begin{equation}
n_b = \gamma_b \int \frac{d^3 k}{(2\pi)^3}  \left[f_{b\,+} - f_{b\,-}  \right]
\end{equation}
where  $\gamma_b=2J_b+1 =2$ is the spin degeneracy factor for the baryon octet, with $J$ the total angular momentum. Moreover, $f(k)$ is the Fermi--Dirac distribution function
\begin{equation}
    f_{b \pm}(k) = \frac{1}{1+\exp[(E_b \mp \mu^\ast_b)/T]} \nonumber
\end{equation}
with energy $E_b= \sqrt{k^2 + {m_b^\ast}^2}$.
Interchanging $b\leftrightarrow d$ the degeneracy factor of the $\Delta$-resonances becomes $\gamma_d=2J_d+1 =4$ and
$E_d= \sqrt{k^2 + {m_d^\ast}^2}$. 
The effective chemical potentials read
\begin{align}
    \mu_{b,d}^\ast &= \mu_{b,d}- g_{\omega {b,d}} \omega_0 - g_{\rho {b,d}} I_{3{b,d}} \rho_{03} - g_{\phi {b}} \phi_0 - \Sigma^r,
\end{align}
where $\Sigma^r$ is the rearrangement term due to the density-dependent couplings
\begin{align}
    \Sigma^r ={}& \sum_b \Bigg[ \frac{\partial g_{\omega b}}{\partial n_b} \omega_0 n_b + \frac{\partial g_{\rho b}}{\partial n_b} \rho_{03} I_{3b}  n_b+ \frac{\partial g_{\phi b}}{\partial n_b} \phi_0 n_b \nonumber \\
    &- \frac{\partial g_{\sigma b}}{\partial n_b} \sigma_0 n_b^s + b\leftrightarrow d\Bigg].
\end{align}
The effective masses are 
\begin{equation}
    m_b^\ast =m_b- g_{\sigma b} \sigma_0, \quad\quad m_d^\ast =m_d- g_{\sigma d} \sigma_0,
\end{equation}
and the scalar density
\begin{equation}
    n_{b}^s =\gamma_b \int \frac{d^3 k}{(2\pi)^3} \frac{m^\ast_b}{E_b} \left[f_{b\,+} + f_{b\,-}  \right].
\end{equation}
We obtain equivalent expressions above for the $\Delta$-resonances by replacing $b$ with $d$. The mesonic mean-field approximation yields
\begin{align}
    &m^2_\sigma \sigma_0 = \sum_b g_{\sigma b}n^s_b + \sum_d g_{\sigma d}n_d^s,\\
    &m^2_\omega\omega_0 = \sum_b g_{\omega b}n_b + \sum_d g_{\omega d}n_d,\\
    &m^2_\phi \phi_0 = \sum_b g_{\phi b}n_b, \\
    &m^2_\rho\rho_{03} = \sum_b g_{\rho b}n_bI_{3b} + \sum_d g_{\rho d}n_dI_{3d}.
\end{align}
The baryon energy and pressure are given by
\begin{flalign}\label{1a}
    \varepsilon_B&=  \varepsilon_b + \varepsilon_m + \varepsilon_d +\varepsilon_L\\
    P_B&=  P_b + P_m +P_d + P_L + P_r\label{1b}
\end{flalign}
with the baryonic contributions
\begin{equation}\label{eq:ener_b}
    \varepsilon_b= \gamma_b \int \frac{d^3 k}{( 2\pi)^3} E_b \left [f_{b+} +f_{b-} \right],
\end{equation}
%
\begin{equation}\label{eq:press_b}
    P_b= \gamma_b \int \frac{d^3 k}{( 2\pi)^3} \frac{k}{E_b} \left [ f_{b+} +f_{b-} \right],
\end{equation}
and the meson contributions
\begin{equation}\label{eq:ener_m}
    \varepsilon_m=  \frac{m_\sigma^2}{2} \sigma_0^2+\frac{m_\omega^2}{2} \omega_0^2 +\frac{m_\phi^2}{2} \phi_0^2  + \frac{m_\rho^2}{2} \rho_{03}^2 ,
\end{equation}
\begin{equation}\label{eq:press_m}
    P_m= -  \frac{m_\sigma^2}{2} \sigma_0^2 +\frac{m_\omega^2}{2} \omega_0^2 +\frac{m_\phi^2}{2} \phi_0^2  + \frac{m_\rho^2}{2} \rho_{03}^2.
\end{equation}
The expressions for $P_d$ and $\varepsilon_d$ are similar to (\ref{1a}) and (\ref{1b}) with the replacement of $b$ with $d$. The pressure further receives a correction from the rearrangement term to guarantee thermodynamic consistency and energy-momentum conservation~\citep{Typel1999, PhysRevC.52.3043}
\begin{equation}
    P_r = n_B \Sigma^r.
\end{equation}
The free Fermi gas contribution of the leptons are accounted for in $\varepsilon_\lambda$ and $P_\lambda$.


From these quantities, we can finally calculate the baryon-free energy density 
${\cal F}_B
= \varepsilon_B - T s_B$, and the 
entropy density
\begin{equation}
     s_B = \frac{\varepsilon_B +P_B- 
     \sum_b \mu_b n_b -
     \sum_d \mu_d n_d 
}{T}.
\end{equation}

\subsection{The Equilibrium Conditions}\label{NS3}

We implement numerical codes to solve the equations of motion for the meson fields, scalar, and baryon densities, and temperature profile by fixing $S/n_B$ and $Y_{L,e}$ towards the study of PNSs. {Here, $Y_{L,e} = (n_e + n_{\nu e})/n_B$, where $n_e$ and $n_{\nu e}$ are the electron and electron neutrino number densities respectively.} A newly born PNS contains trapped neutrinos, so it is standard to consider the electron and the muon lepton numbers as fixed. In our calculations in the neutrino-trapped regime, we fix the electron lepton number $Y_{L,e}=Y_e + Y_{\nu e}$ and ignore the contribution of the muon and muon neutrino {$Y_{L,\mu}= Y_\mu +Y_{\nu_\mu}=(n_\mu + n_{\nu \mu})/n_B \approx 0$ in accordance with supernova physics~\citep{Malfatti}, with $n_\mu$ and $n_{\nu \mu}$ being the muon and muon neutrino number densities respectively}. We consider different values of $S/n_B$ for different $Y_{L,e}$ in accordance with the various stages of PNS evolution~\citep{Nakazato, Raduta1}: for the newly born neutron star (at $t=0\,\text{s}$) we consider  $S/n_B =1$ and $Y_{L,e} = 0.4$, but a few seconds ($\sim\,0.5-1.0 ~\text{s}$) after the star is born it starts heating, so the entropy increases ($1\,<\, S/n_B\,<3$) and the lepton number concentration decreases, thus we consider $S/n_B =2$ and $Y_{L,e} = 0.2$ at this stage.  Further discussions on fixed entropy calculations can be found in~\citet{Raduta}. The star gets maximally heated and becomes neutrino-free ($Y_{\nu e}=0$) with $S/n_B = 2$, and finally, it shrinks to a catalyzed cold neutron star at $T= 0$,~\citep{Steiner, Shao, Reddy}.  For the neutrino-free region, we consider both electrons and muons in the calculation.  A snapshot of each stage is discussed in detail below in Fig.~\ref{NHD1}. 

We consider the matter to be in $\beta$-equilibrium during all the stages, and use the following relations for the chemical potentials:
\begin{align}
    &\mu_\Lambda = \mu_{\Sigma^0} = \mu_{\Xi^0} = \mu_{\Delta^0} = \mu_{n}=\mu_B,\\
    &\mu_{\Sigma^-} = \mu_{\Xi^-} = \mu_{\Delta^-} = \mu_{B}-\mu_Q,\\
    &\mu_{\Sigma^+} = \mu_{\Delta^+} = \mu_p=  \mu_{B}+\mu_Q, \\
    &\mu_{\Delta^{++}}=\mu_{B}+2\mu_Q,
\end{align}
with $\mu_B$ the baryon chemical potential and $\mu_Q=\mu_p-\mu_n$ the charged chemical potential. 

In the neutrino-trapped region, the charge chemical potential can be expressed in terms of the lepton and neutrino chemical potentials as
\begin{equation}
  \mu_Q= \mu_{\nu l}-\mu_l , 
\end{equation}
where $l$ is a lepton i.e., either electron $e$ or muon $\mu$ and $\mu_{\nu l}$ is the neutrino chemical potential. In the neutrino transparent region, the chemical potential of the neutrinos vanishes and the lepton chemical potential is related to the charge chemical potential as 
\begin{equation}
    \mu_Q =-\mu_l.
\end{equation}
Also, lepton number densities are conserved, $Y_{L,l}=Y_l+Y_{\nu l}$ in the neutrino-trapped matter. The system is charge neutral, so baryon and lepton charges must cancel out
  \begin{equation}
  \begin{split}
n_p+&n_{\Sigma^+}+2n_{\Delta^{++}}+n_{\Delta^+}\\
& -(n_{\Sigma^- }+ n_{\Xi^-} + n_{\Xi^-} + n_{\Delta^-})=  n_e + n_\mu.
\end{split}
\end{equation}




\section{Results and Analysis}\label{RA}

\begin{figure*}
  \centering
  \includegraphics[scale=0.55]{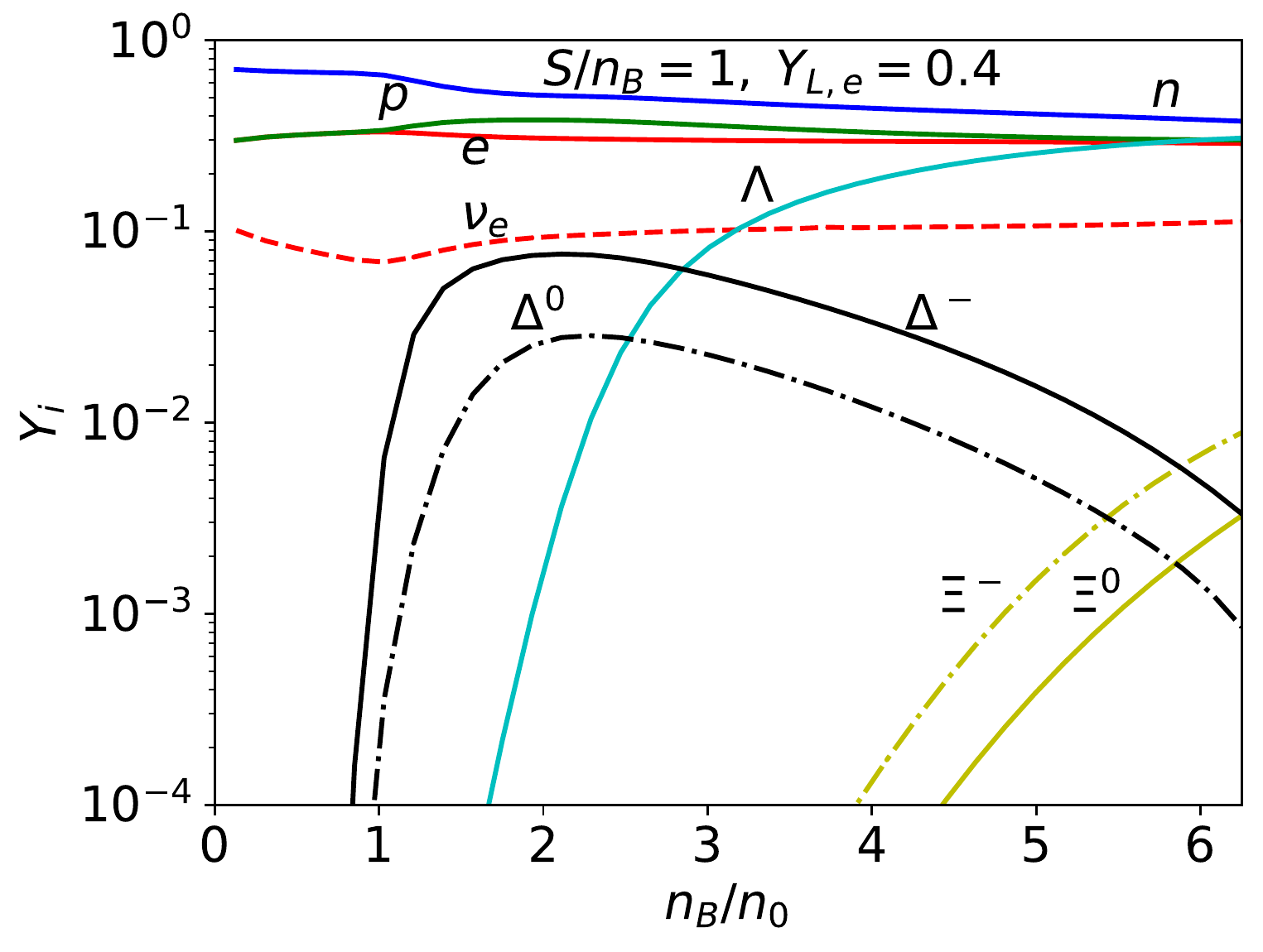}
  \qquad
  \includegraphics[scale=0.55]{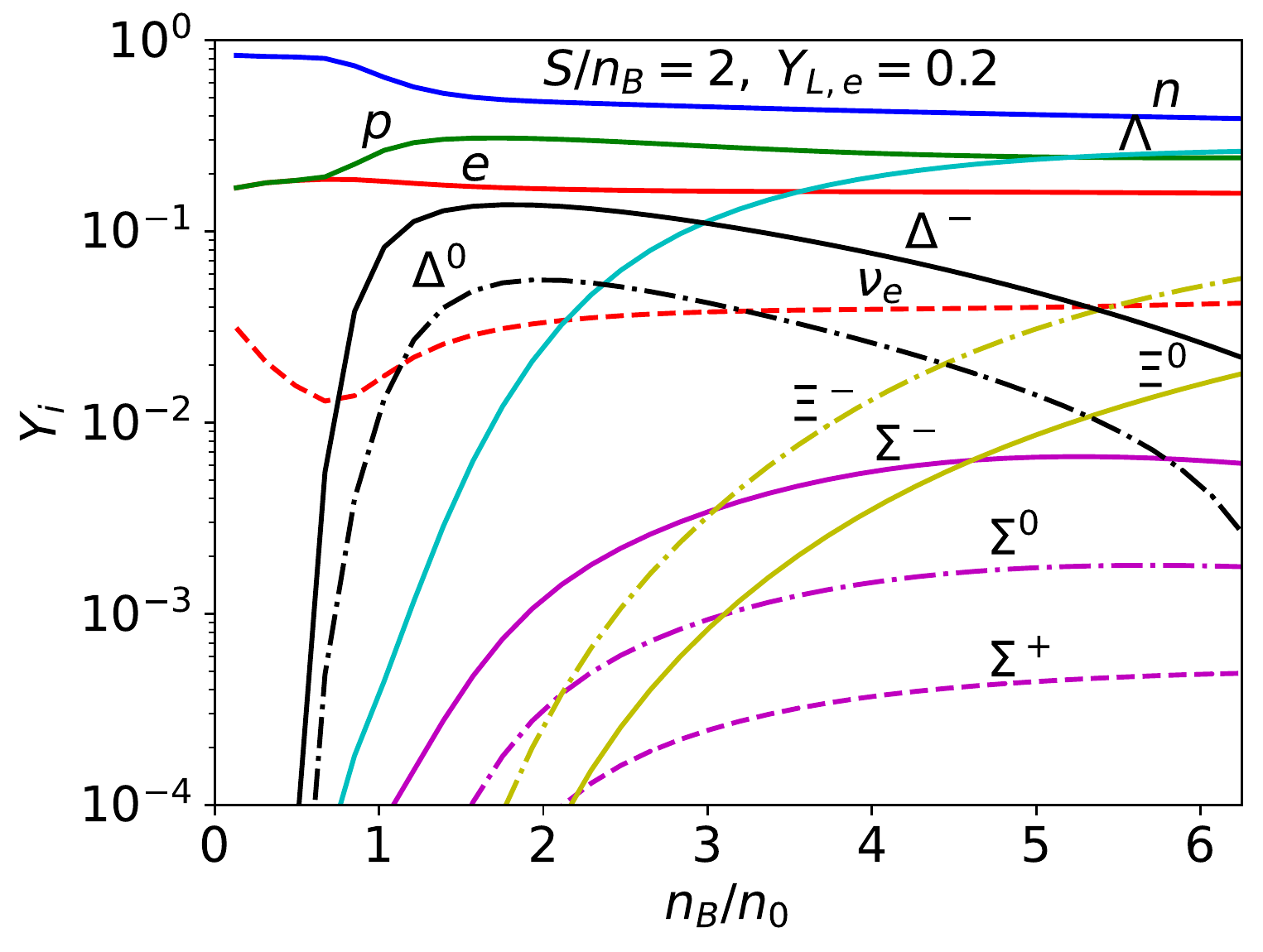}
   \qquad
  \includegraphics[scale=0.55]{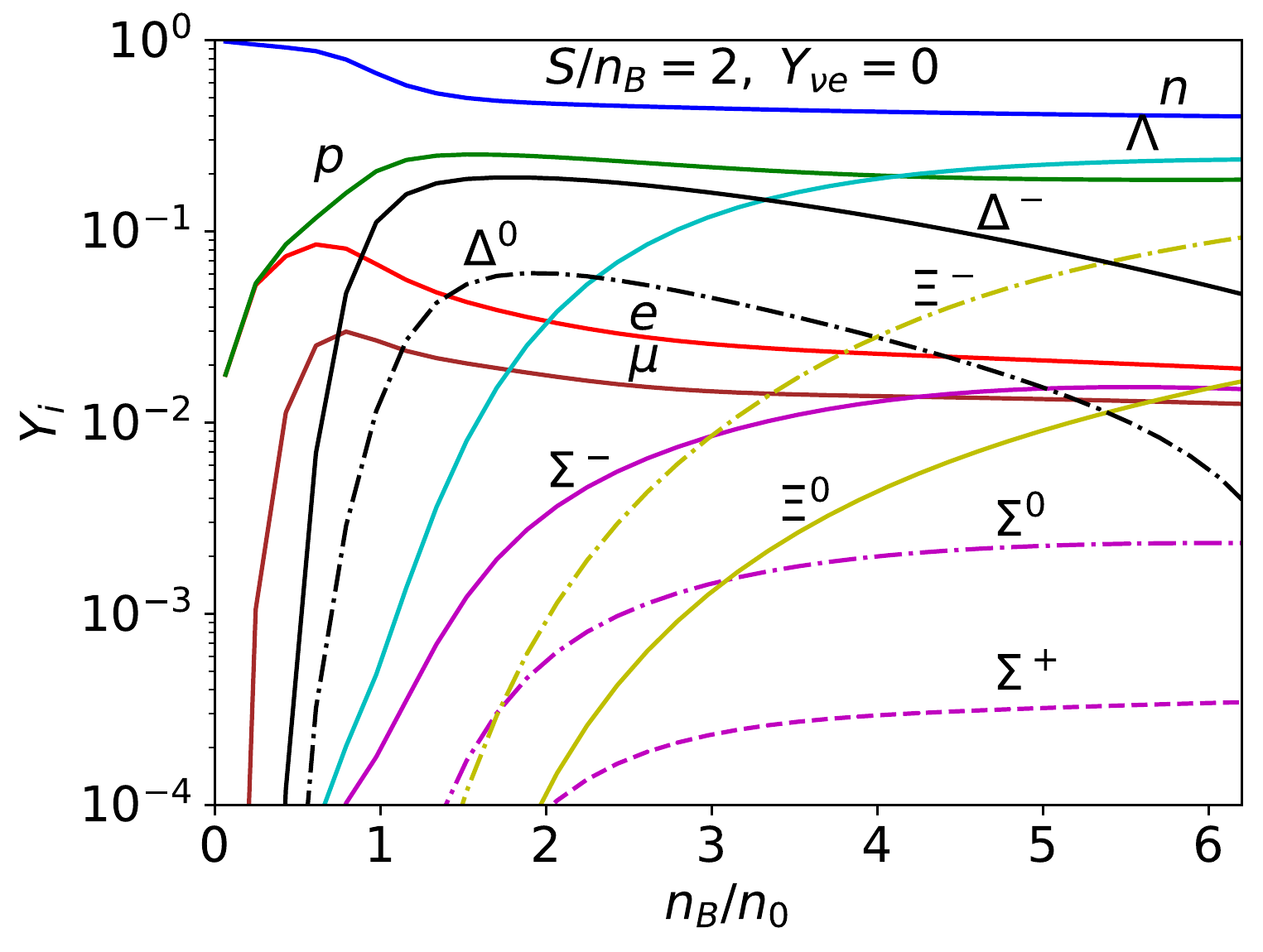}
   \qquad
    \includegraphics[scale=0.55]{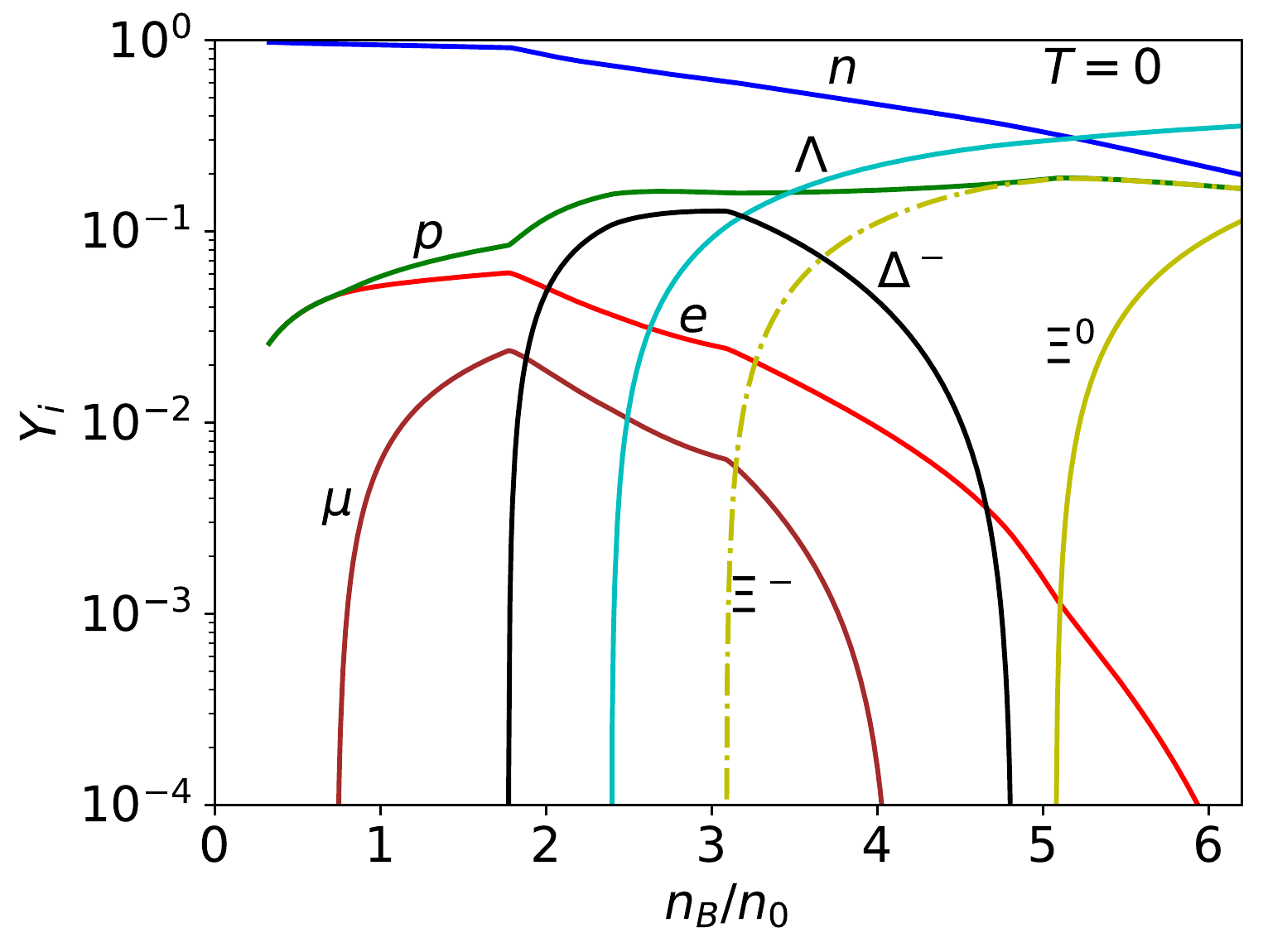}
   
   \caption{
   The figures show the particle fraction ($Y_i$) as a function of baryon density at various stages of PNS evolution. The upper panels show the results for the neutrino-trapped region; the first panel from the left $S/n_B =1$, $Y_{L,e}=0.4$ shows when the star was born, and the second panel $S/n_B =2$, $Y_{L,e}=0.2$ shows when the star starts deleptonization following neutrino diffusion. The lower panels show the neutrino-free regime of the star; the first panel $S/n_B = 2$, $Y_{\nu e} = 0$ shows a maximally heated star and the right panel $T= 0$ shows the stage where a cold {\it neutron star} is born. 
   }
   \label{NHD1}
\end{figure*}

In Fig.~\ref{NHD1} 
we present the results for particle composition of nucleonic ($N$), hyperonic ($H$), and $\Delta$-resonances ($\Delta$) admixed hypernuclear matter in PNS core during its evolution till it catalyzes into a {\it neutron star}. {The quantity $Y_i$ is the particle fraction which can be expressed explicitly as $Y_i = n_i/n_B$, where $i$ represents the different particles in the system.} In the upper panels from left to right, {we observe that the ratio of proton fraction ($Y_p$) to the neutron fraction ($Y_n$), $Y_p/Y_n$, decreases across the panels.
However, the asymmetry ($\delta$) between proton and neutron in the system is given by $\delta = (n_n-n_p)/n$, with $n = n_n + n_p$, while $n_n$ and $n_p$ are neutron and proton number densities respectively. Therefore, a decrease in $Y_p/Y_n$ results in larger values of $\delta$ making the system  even more asymmetric across the panels from left to right.} This is also true for the neutrino-free regime in the lower panels from left to right. Comparing the particle fractions for the evolution of the star in the neutrino-trapped matter; we observe two main effects, the abundance of the neutrinos affects the $Y_p/Y_n$, and the appearance of particles at low baryon densities. Trapped neutrinos delay the appearance of heavy baryons in general, at low densities, and further delay the appearance of strange matter constituents to higher densities. Comparing the upper panels (ambient condition of core birth at various stages) and the lower panels (ambient conditions after deleptonization at various stages) we observe that neutrino-trapping increases proton and electron concentration in the stellar matter. At $S/n_B =1$, $Y_{L,e} = 0.4$, the $\Delta$-resonances start appearing at densities equivalent to the saturation density, {firstly the  $\Delta^- $, and then the $\Delta^0,$} before the appearance of the first particle with strangeness, the $\Lambda$. Subsequently, heavy baryons start appearing at relatively low densities during deleptonization: at densities lower than the saturation density. When $S/n_B = 2$, $Y_{L,e}=0.2$ the baryonic composition of the star up to $n_B\sim\,2n_{0}$ is {$\Delta^-,\,\Delta^0,\, \Lambda,\,\Sigma^-,\,\Sigma^0\;\text{and}\; \Xi^-$}. We can infer that during the early stages of PNS evolution, the stellar matter is mostly composed of non-strange baryons while strange matter constituents are found at higher densities, towards the center of the star. 

\begin{figure}
  \centering
  \includegraphics[scale=0.6]{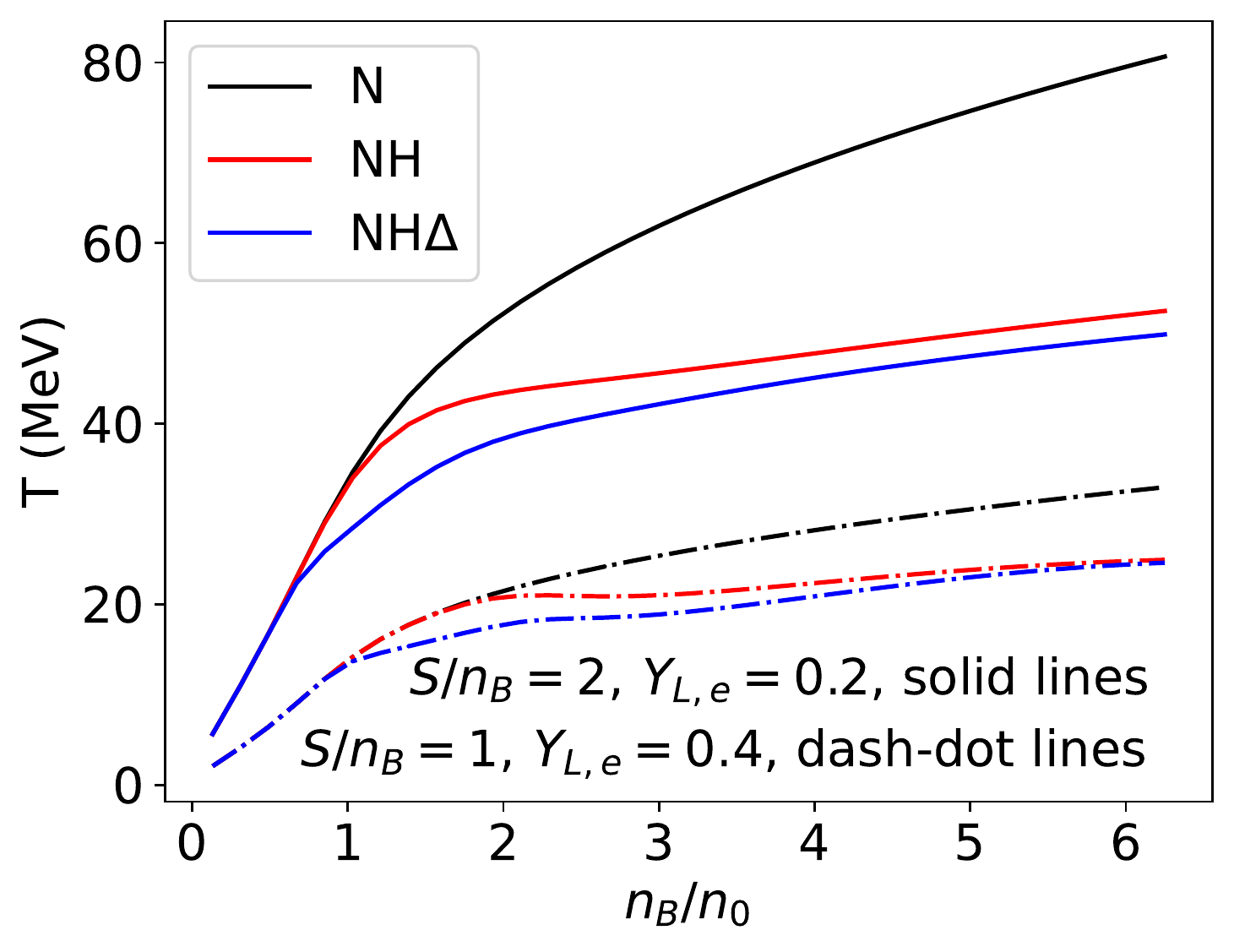}
  \includegraphics[scale=0.6]{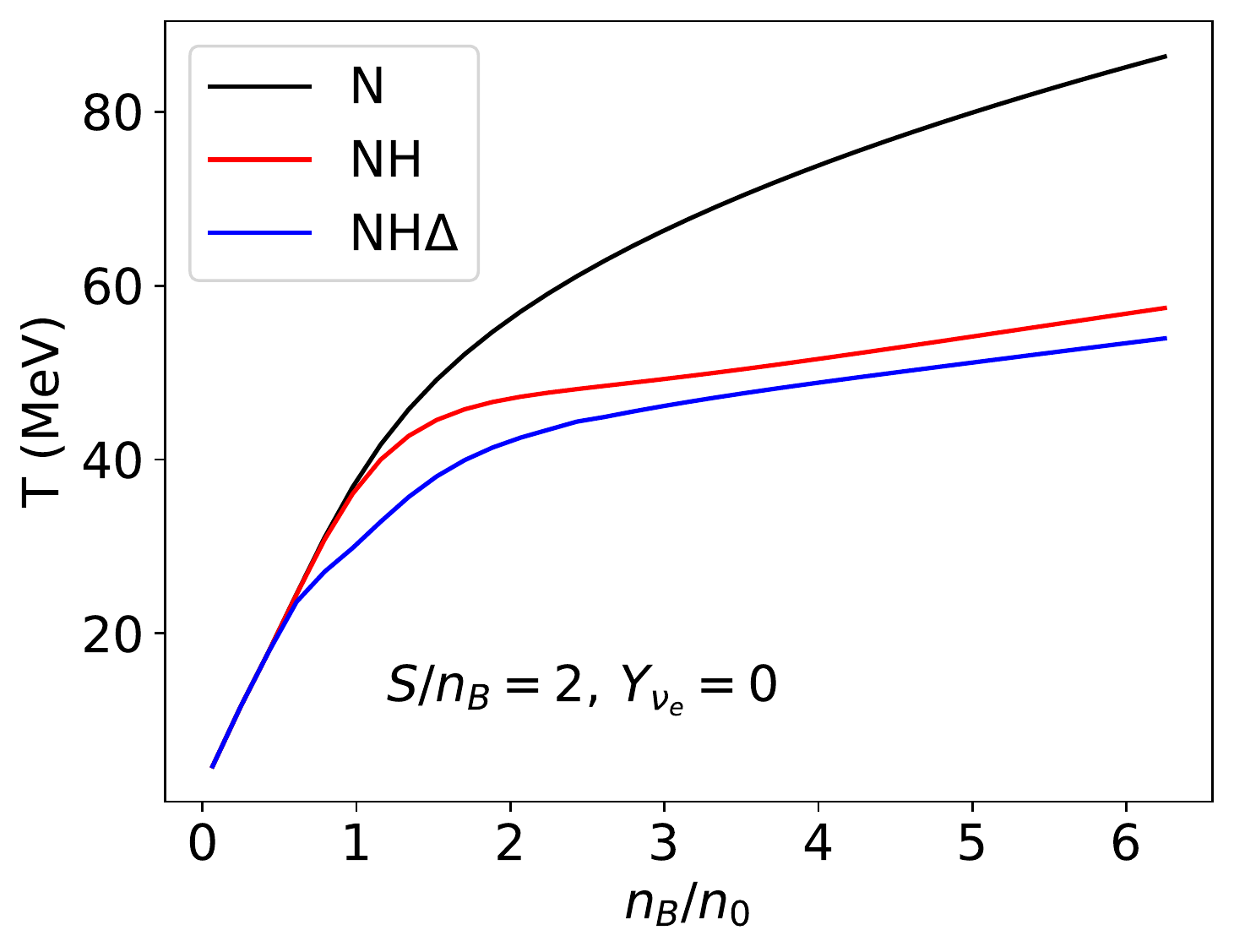}
   \caption{The figures show the temperature profiles in a PNS at different stages of its evolution. The upper panel shows the temperature profiles of the neutrino-trapped region and the lower panel shows the temperature profiles of the neutrino-free region of the evolution.}
    \label{TNHD}
\end{figure}

Additionally, in the neutrino-free phase of the star's evolution, the bottom panels from left to right, the strange matter population at {densities lower than $2n_{0}$ decreases} as the star cools down. The heavy baryon content of the stellar matter at $n_B\sim 2n_{0}$, when it cools down to $T = 0$, is $\Delta^-$. The $\Lambda$ appears slightly further from $2n_{0}$, and the $\Xi^-$ appears around $3n_{0}$. In sum, the strange matter constituents are suppressed to higher densities when the entropy in the core is low~\citep{Prakash1}. {In general, the threshold density for the emergence of the hyperons decreases with increasing entropy and decreasing lepton number density. This implies higher temperatures favor the appearance of the hyperons at low densities since an increase in entropy is accompanied by an increase in temperature --- see Fig.~\ref{TNHD}.} In the bottom panel with $S/n_B=2$, the star is lepton-poor but still hot, {indeed}, the star is expected to reach its maximum temperature when $S/n_B = 2$, $Y_{\nu e}=0$ before it starts cooling. This can be seen in Fig.~\ref{TNHD} below. {The star then continues cooling until it forms a cold \textit{neutron star} at $T=0$. }

\begin{figure}
  \centering
  \includegraphics[scale=0.55]{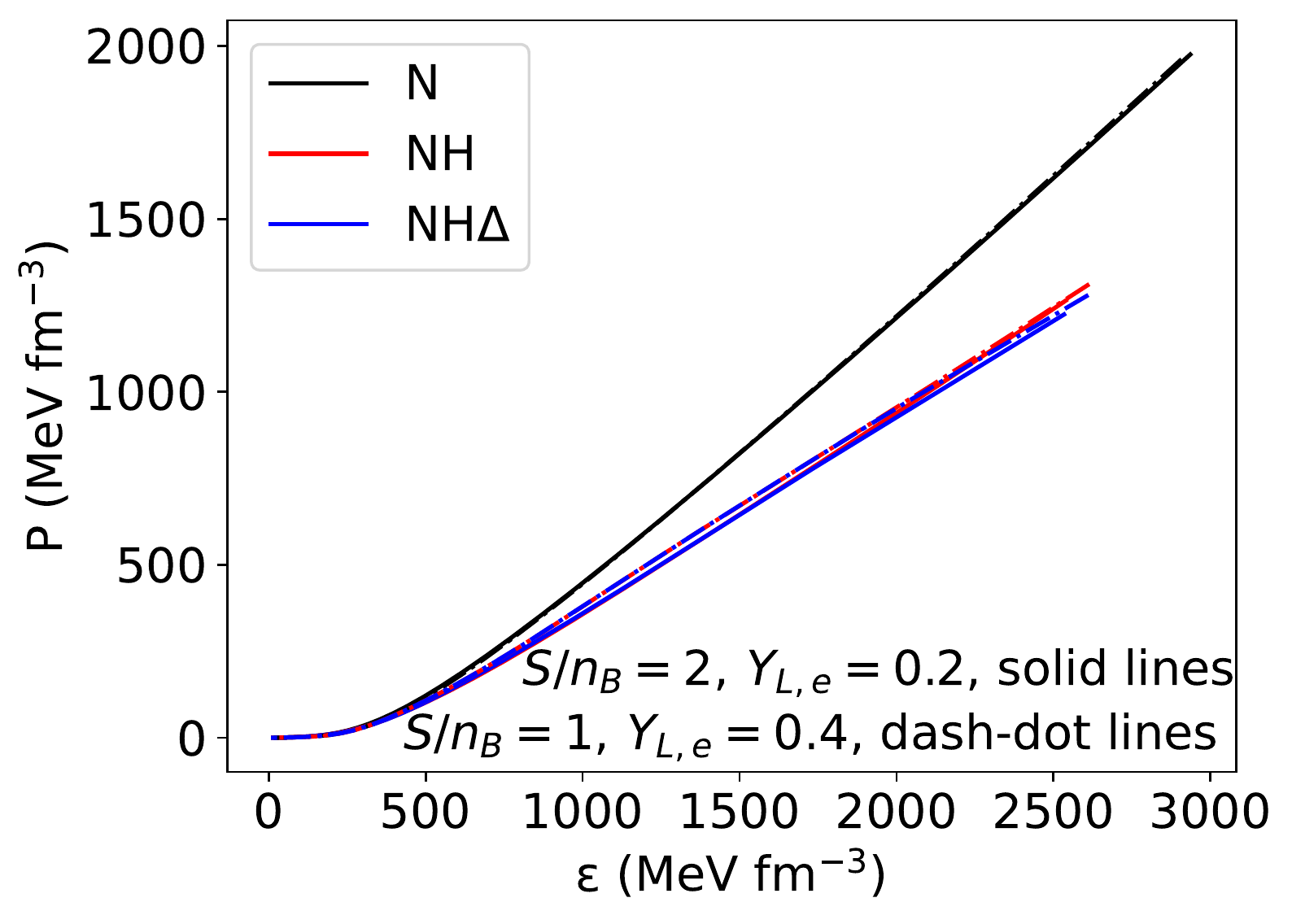}
  \qquad
  \includegraphics[scale=0.55]{ 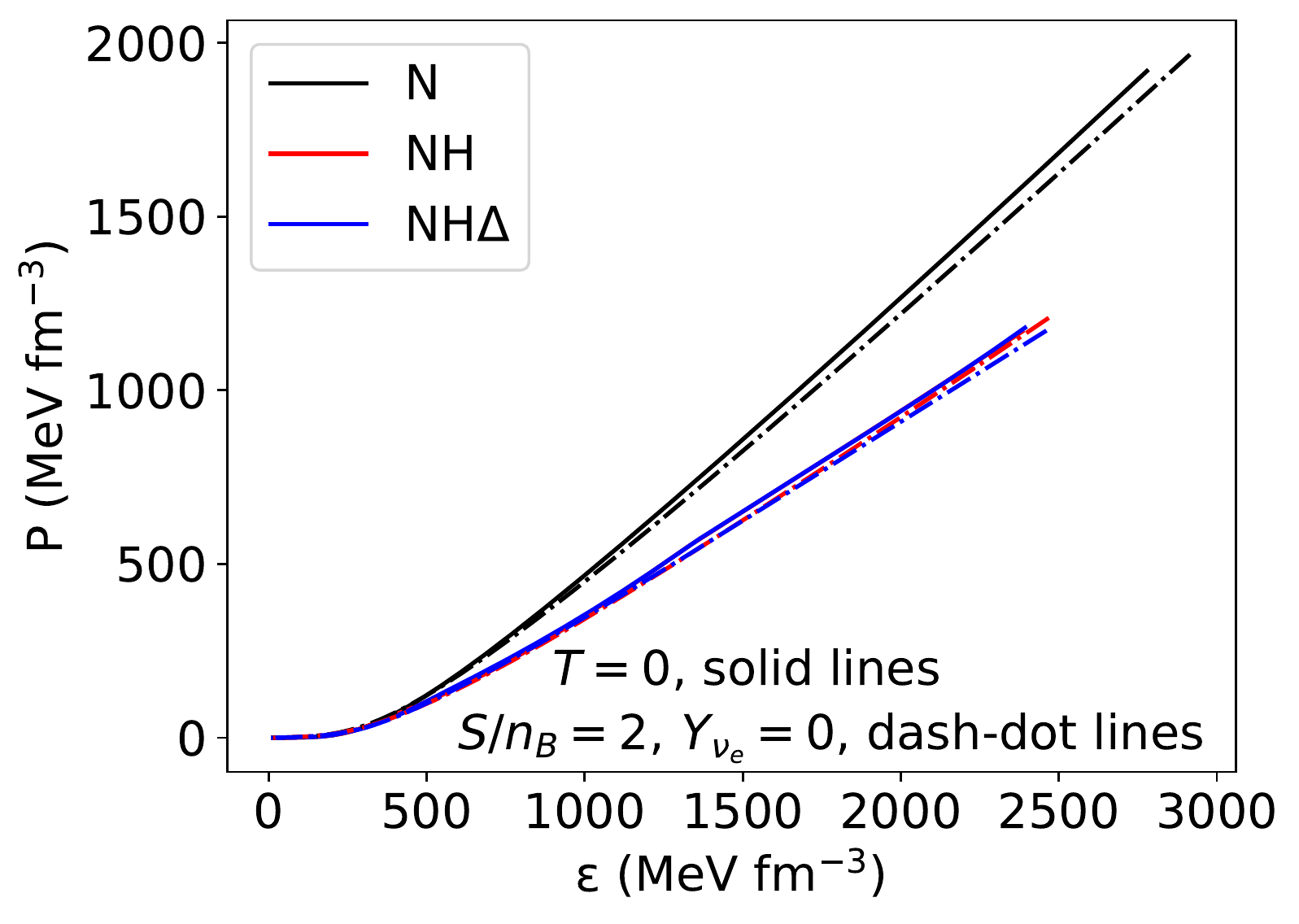}
   
   \caption{We present the EoS composed of $N$, $NH$, and $NH\Delta$ at various stages of PNS evolution. The upper panel shows the EoS for neutrino-trapped matter for different $S/n_B$ and $Y_{L,e}$, and the lower panel shows the EoS for the neutrino-free region for $S/n_B=2$ and $T=0$.}
   \label{EoS}
\end{figure}

In Fig.~\ref{EoS} we present the results for the EoS of hot star matter at various stages of evolution for $N$, $H$, and $\Delta$-resonances admixed hypernuclear matter in $\beta$-equilibrium at a fixed entropy. We show the results for the pressure $P$ as a function of the total energy density $\varepsilon$. The figure in the top panel represents the EoS for neutrino-trapped matter and the bottom panel represents neutrino-free matter. The EoS for $N$, $NH$, and $NH\Delta$ hypernuclear matter becomes stiffer with decreasing $Y_{L,e}$ ($0\,\leq Y_{L,e}\,\leq \,0.4$) and $S/n_B$ ($1\,\leq S/n_B\,\leq\,2$).This behavior is accompanied by a decrease in $Y_p/Y_n$ as the primary attribute {which makes the system more asymmetric thereby increasing the symmetry energy.} The inclusion of hyperons to hypernuclear matter generally softens the EoS while the $\Delta$-resonances soften the EoS at low to intermediate densities and stiffen it at higher densities. This observation is well established in zero temperature studies of neutron stars~\citep{Schurhoff, Drago1, Cai, Ribes, Sahoo}. We see from Fig.~\ref{NHD1} that the presence of a large electron neutrino fraction delays the appearance of the hyperons to higher densities while the low electron content at higher entropies enhances the appearance of the heavy baryons at low densities. The appearance of heavy baryons at low densities significantly softens the EoS, both for neutrino-free and neutrino-trapped stellar matter, which is a well-known result also related to the hyperon puzzle \citep{Menezes}

In Fig.~\ref{TNHD}, we present the results for temperature as a function of baryon density for hot hypernuclear matter composed of nucleons, nucleons, and hyperons, and nucleons, hyperons and $\Delta$-resonances. The diagram in the upper panel is the temperature profile in neutrino-trapped hot star for different $S/n_B$ and $Y_{L,e}$ while the lower panel represents the temperature profile for neutrino transparent matter for $S/n_B = 2$. 

Generally, the changes in the slope of the figures are attributed to the appearance of heavy baryons. For pure nucleonic matter, the temperature increases steadily with baryon density. {When hyperons and $\Delta$-resonances are introduced into the hypernuclear matter they decrease the temperature significantly, and we start seeing a departure from the $N,\, NH,\,\text{and}\; NH\Delta$ curves at baryon densities between $n_{0},\,\text{and}\; 2n_{0}$. It is worth mentioning that introducing additional particles into the system increases its entropy per baryon, 
which is accounted for by the temperature drop in the system. Thus, the temperature profile for nucleon-only stellar matter is higher than that of the nucleon plus hyperon admixture which is, in turn, higher than nucleon plus hyperon plus $\Delta$- resonances admixture.}  This observation is in agreement with the discussions in~\citet{Oertel, Raduta1} which argue that the entropy of a hypernuclear system increases with the number of constituent particles. In that regard, in a system with fixed entropy, an increase in constituent particles leads to an increase in the specific heat of the system which favors a temperature decrease. That notwithstanding, it has been argued in~\citet{Mayle} that the introduction of negatively charged particles into hypernuclear matter other than electrons reduces the net electron number density, releasing electron degeneracy energy and resulting in a high-temperature supernova core. 

Comparing the temperature profiles to the particle abundances, we observe that the temperature profile for $N$ and $NH\Delta$ start departing from each other at a density in which the first $\Delta$-resonance baryon, i.e. when $\Delta^-$ appears in the matter for each system. Likewise, the temperature profile for $NH$ departs from $N$ at a density in which the first strange particle appears in the matter, mostly, the $\Lambda$-particle. Moreover, hyperons and $\Delta$-resonances appear at relatively lower densities for higher entropy, $S/n_B = 2$ matter as in Fig.~\ref{NHD1}, this reflects in the temperature profiles as well. The characteristics of the temperature profile are attributed to the appearance of new particles introducing new degrees of freedom  and altering the specific heat of the system which is compensated by the change in temperature to keep the entropy fixed --- see~\citet{Sedrakian, Raduta1} for more discussion. 




\begin{table}
\begin{center}

\begin{tabular}{l|l|l|l}
      $S/n_B;\,Y_{L,e}$             & Matter content                                                                 & $M_{\text{max}}/M_{\odot}$                                                                    & $R/km$                                                                  \\ \hline
$1;\; 0.4$ & \begin{tabular}[c]{@{}l@{}}$N$\\ $NH$ \\ $NH\Delta$\end{tabular}  & \begin{tabular}[c]{@{}l@{}}$2.44$\\ $2.32$\\ $2.32$\end{tabular} & \begin{tabular}[c]{@{}l@{}}$12.34$\\$12.41$\\ $12.41$\end{tabular} \\ \hline
$2;\; 0.2$ & \begin{tabular}[c]{@{}l@{}}$N$\\ $NH$ \\ $NH\Delta$\end{tabular} & \begin{tabular}[c]{@{}l@{}}$2.49$\\$2.29$\\ $2.29$\end{tabular} & \begin{tabular}[c]{@{}l@{}}$12.83$\\ $12.59$\\ $12.56$\end{tabular} \\ \hline
$2;\; Y_{\nu e}=0$ & \begin{tabular}[c]{@{}l@{}}$N$\\ $NH$ \\ $NH\Delta$\end{tabular}  & \begin{tabular}[c]{@{}l@{}}$2.49$\\ $2.24$\\ $2.24$\end{tabular} & \begin{tabular}[c]{@{}l@{}}$12.87$\\$12.51$\\ $12.41$\end{tabular}\\ \hline
$ T=0$ & \begin{tabular}[c]{@{}l@{}}$N$\\ $NH$ \\ $NH\Delta$\end{tabular}  & \begin{tabular}[c]{@{}l@{}}$2.48$\\ $2.26$\\ $2.26$\end{tabular} & \begin{tabular}[c]{@{}l@{}}$12.03$\\$11.96$\\ $11.91$\end{tabular}\\
\hline
\end{tabular}
\caption{Maximum masses ($M_{\text{max}}$) and radii ($R$) of stellar matter}
\label{ma}
\end{center}
\end{table}



\begin{figure}
    \centering
    \includegraphics[scale=0.7]{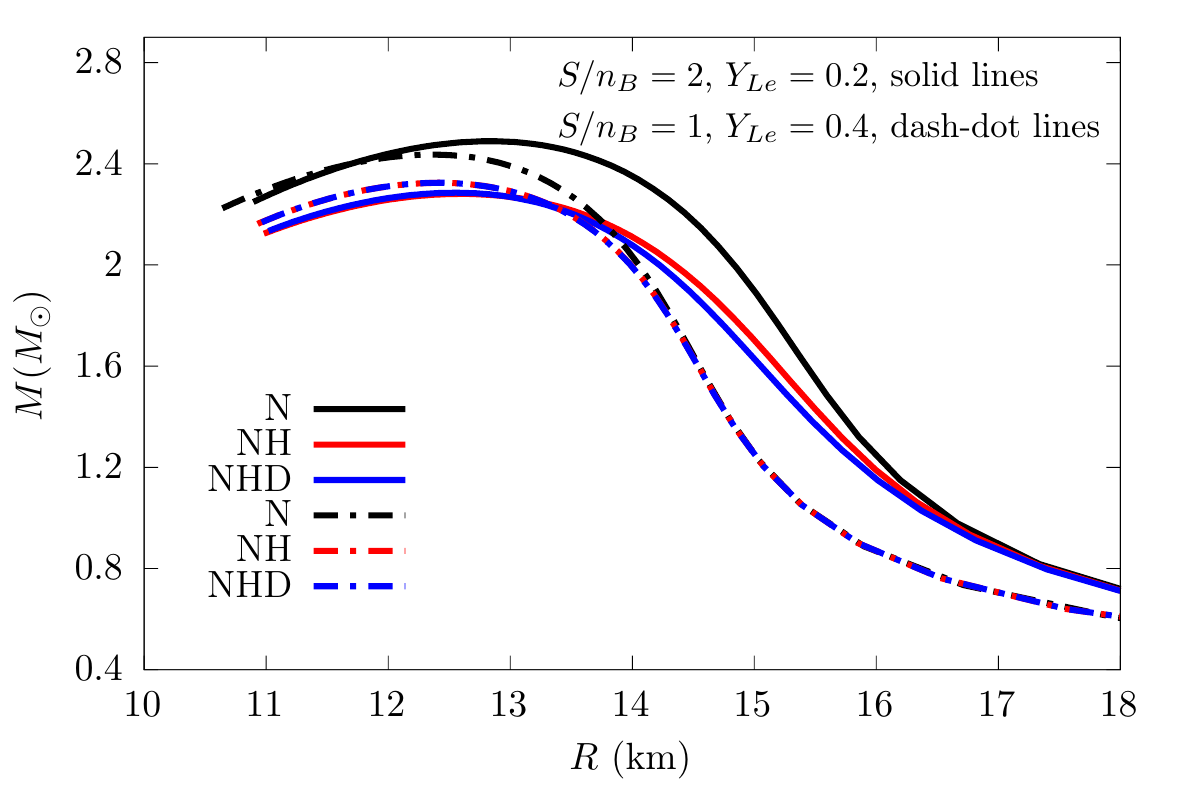}
    \includegraphics[scale=0.7]{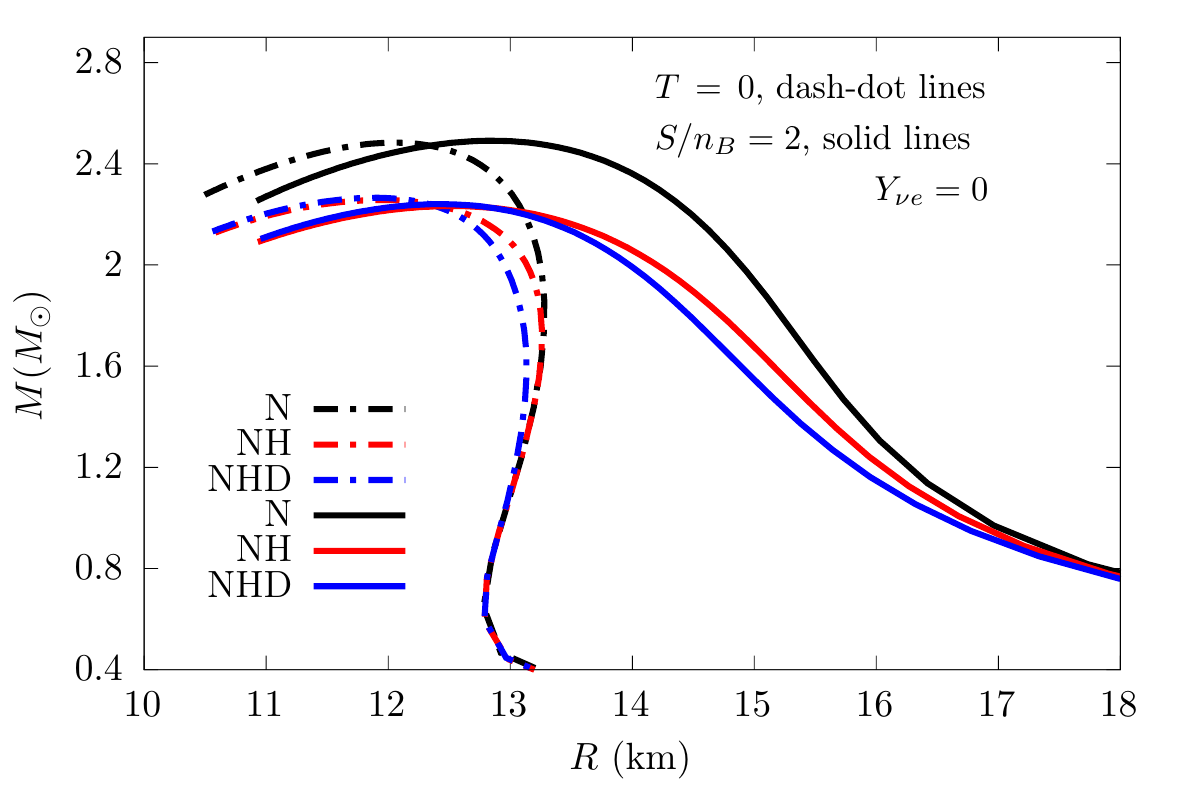}    
   
   \caption{Gravitational mass $M$ of a PNS as a function of radii $R$ for non-rotating spherically-symmetric stars. The top panel shows the results for neutrino-trapped $\beta$-equilibrated star matter at different stages of the star's evolution with different $S/n_B$ and $Y_{L,e}$. The bottom panel shows a neutrino-transparent star for $S/n_B=2$ and $T=0$.}
   \label{mr}
\end{figure}


In Fig.~\ref{mr} we show the results of the gravitational mass of stars as a function of their radii at different stages of their evolution for baryonic matter composed of $N,\, NH,\,\text{and}\, NH\Delta$. The onset of new degrees of freedom is distinctively represented by different curves with different slopes. The top panel shows the regime in which the neutrinos are trapped inside the star at different $S/n_B$ and $Y_{L,e}$ while the bottom panel shows the results for neutrino transparent region of the star for $S/n_B = 2$ and $T=0$. Generally, the presence of hyperons and $\Delta$s are expected to reduce the maximum mass of the star, preventing it from reaching the maximum observable mass~\citep{Antoniadis, Demorest}. One way of dealing with this problem is through a consistent definition of the baryon-meson coupling. That notwithstanding, the model under investigation is compatible with the $2M_{\odot}$ constraint. We observe from Tab.~\ref{ma} and the figures that the radius of the star increase with increasing $S/n_B$ and decreasing $Y_{L,e}$. This is because at higher entropies the star gets heated and expands and its mass also reduces due to neutrino diffusion. 

Aside from the discussions above, we employ different couplings and carry out the study at fixed lepton number density and entropy. This makes our results for $S/n_B= 1,\;\text{and}\;2$ different from cold $\beta$-equilibrated neutron stars. As can be observed in Fig.~\ref{mr}, the mass-radius diagrams for hot non-rotating spherically symmetric neutron stars, the intermediate-masses, and the maximum masses presented in Table.~\ref{ma} have radii relatively large for both neutrino-trapped and neutrino-transparent matter compared to a cold \textit{neutron star}. This is attributed to the hot nature of the stars under study in those stages. 

\section{Conclusions}\label{Con}
We investigated the presence of exotic baryon contents in neutron stars from birth through a supernova explosion until it catalyzes to form a cold neutron star. A relativistic model within a mean-field approximation was used for this work. The snapshots of the particle abundances at various stages of the star evolution are presented in Fig.~\ref{NHD1}. We examined the EoS for $N$, $NH$, and $NH\Delta$ mater and observed that $NH$ and $NH\Delta$, soften the EoS, as is well known, and the results are presented in Fig.~\ref{EoS}. The temperature profiles during the evolution of the star were also studied. The inclusion of new particles, such as the hyperons, reduces the temperature below the nucleon-only stellar matter and the further addition of $\Delta$-resonances to nucleon plus hyperons further decreases the temperature of the stellar matter. Consequently, the presence of hyperons and $\Delta$-resonances increases the specific heat, leading to a decrease in the temperature gradient. The mass-radius diagram was also studied and the results are presented in Fig.~\ref{mr}. The evolution stages of the star are summarized below:

\begin{itemize}
    \item First stage: $S/n_B = 1$, $Y_{L,e}=0.4$, this is a neutrino-trapped regime at the early stages of the evolution. Here, the heavy baryons appear at densities greater than the saturation density, $n_B\,>\,n_0$. The particle content up to $n_B\sim 2n_0$ is in the order $\Delta^-,\;\Delta^0,\;\text{and}\;\Lambda$. The temperature profile of the stellar matter at this stage is relatively less than the neutrino diffusion stage and it has a relatively stiffer EoS and smaller radii.
    
    \item Second stage: $S/n_B = 2$, $Y_{L,e}=0.2$, this is the deleptonization stage where the star gets heated and expands due to neutrino diffusion. The temperature profile at this stage is higher than in the first stage and the EoS softens with relatively high star radii. At this stage, the heavy baryons shift more towards lower baryon densities (less than the saturation density), and the order of particle appearance up to $n_B\sim 2n_{0}$ are $\Delta^-,\; \Delta^0,\;\Lambda,\;\Sigma^-,\, \Sigma^0,\,\text{and}\;\Xi^-$. Thus, the neutrino abundance suppresses the appearance of the heavy baryons and delays the strange matter particles to higher baryon densities, comparing this and the  first stage.

    \item Third stage: $S/n_B = 2$, $Y_{\nu e}=0$, here the star is maximally heated, neutrino-transparent, and cooling through the emission of pairs of neutrinos. The temperature of the stellar matter here is higher than in the two stages described above, with softer EoS and higher radii. The heavy particles are shifted towards lower baryon densities, in the order $\Delta^-,\;\Delta^0,\;\Lambda,\;\Sigma^-,\;\Sigma^0,\,\Xi^-,\, \Xi^0\,\text{and}\; \Sigma^+$, almost all the particles appear before or at $n_B \,=\, 2n_0$. Here, more strange matter constituents appear at lower densities compared with the previous stage.

    \item Final stage: $T = 0$, at this stage the star is neutrino-transparent and in a catalyzed configuration 
    The star shrinks with stiffer EoS and smaller radii. The heavy baryons shift towards higher baryon densities, $n_B\,>\,n_0$. The heavy baryon content of the stellar matter up to $n_B\sim 2n_0$ is $\Delta^-$. We have one heavy baryon appearing at this density range, at this stage, with it being non-strange. Comparing the three stages above, the heavy baryons shift gradually towards higher densities as the star cools.
\end{itemize}
{The results qualitatively agree with the ones in references \cite{Raduta1, Sedrakian, Malfatti, Sedrakian:2021qjw, Pons} in terms of $\Lambda$ being the most abundant heavy baryon, the softening and hardening of the EoS, the temperature profile of the stellar matter, and the hierarchy of the mass-radius diagram. The evolution stages of the PNS, its structure, and compositions are also discussed in Ref.~\cite{Prakash}.} We observed that the presence of higher temperatures inside the star favors the appearance of heavy baryons at lower baryon densities and vice versa. We can draw a relation between the entropy increase and the softening of the EoS because the appearance of heavy baryons at lower densities means a softer EoS. On average, the most abundant heavy baryons in the star at all the stages of its evolution are; the $\Lambda^0$, which constitutes more than $18\%$ of the matter content followed by $\Delta^-$ which constitutes about $10\%$, followed by $\Xi^-$ (the $\Delta^0$ surpasses it only at the first stage) which constitutes about $7\%$ of the matter content before the next $\Delta$-resonance ($\Delta^0$) forming about $4\%$ of the matter content.

\section*{Acknowledgements}


This work is a part of the project INCT-FNA Proc. No. 464898/2014-5. D.P.M. was partially supported by Conselho Nacional de Desenvolvimento Científico e Tecnológico (CNPq/Brazil) under grant 303490-2021-7. A.I. and K.D.M. were also supported by CNPq/Brazil under grants 168546/2021-3 and 150751/2022-2, respectively. M.R.P. is supported by Conselho Nacional de Desenvolvimento Científico e Tecnológico - Brasil  (CNPq)  and Coordena\c c\~ao de Aperfei\c coamento de Pessoal de N\'ivel Superior (Capes/Brazil) with scholarships.

\section*{Data Availability}
The datasets generated during and/or analyzed during the current study are available from the corresponding author upon reasonable request.

\bibliographystyle{mnras}
\bibliography{new_references}
\label{lastpage}
\end{document}